\documentclass{cflowproc}

\begin{document}


\shortauthors{Arabadjis \& Bautz}        
\shorttitle{Constraining Multiphase Gas} 

\title{Constraining Multiphase Gas in Cooling Flows}   

\author{John S. Arabadjis\affilmark{1}
and Marshall W. Bautz\affilmark{1}}

\affil{1}{Massachusetts Institute of Technology}   


\begin{abstract}
We present a spectral analysis of the central X-ray emission for a sample of
galaxy clusters observed with Chandra.  We constrain the quantity of a second
cospatial temperature component using Markov Chain Monte Carlo sampling and
discuss the implications for our understanding of cooling flows.
\end{abstract}


\section{Introduction}
\label{arabadjis:intro}


The cores of many galaxy clusters are sufficiently dense and cool that the
plasma cooling time is shorter than a Hubble time.  For many years it was
thought that runaway cooling would result in a large central mass deposition
rate \citep{fabian,allen_fabian,peres_etal,white,allen}.  Chandra and XMM
observations have altered this picture significantly -- there appears to
be a core temperature floor of 1-2 keV, and inferred mass deposition rates
have been reduced by an order of magnitude
\citep{peterson_p_etal,peterson_k_etal}.  Several mechanisms have been
proposed to explain the lack of colder gas, including heating by AGN, heat
conduction from cluster halo plasma, and small-scale variations in the cooling
and metallicity structure of the plasma.  Each of these processes can
potentially leave a specific observational signature.  For example, if
conduction provides the energy required to arrest the cooling then the core
plasma may be single-phase, but if AGN and/or small-scale inhomogeneities are
responsible, one might expect to see observational signatures indicating the
presence of multiphase plasma.

We have analyzed a sample of 12 galaxy clusters found in the data archive of
the {\it Chandra X-ray Observatory} for evidence of multiphase plasma in each
cluster core.  We first describe the method, briefly sketching the Markov Chain
Monte Carlo technique, and apply it to a sample of clusters.  We then discuss
the results and their implications for cooling flows.

\section{Method}
\label{arabadjis:method}

\begin{figure}
\plotone{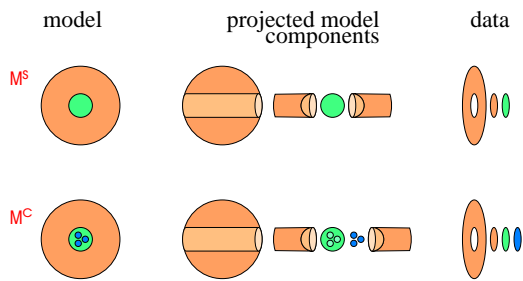}
\figcaption{Core-halo geometry of the simple and complex models.
\label{arabadjis_f1}}
\end{figure}

We use a simple core-halo geometry and Markov Chain Monte Carlo (MCMC)
simulations to assess the statistical significance of the presence of
multiphase plasma in a sample of clusters.  We compare two models of the
emssion: a simple model ${\sf M^s}$ which contains one emission component in
the halo and one in the core, and a complex model ${\sf M^c}$ which contains
one in the halo and two in the core (see Figure~\ref{arabadjis_f1}).

\begin{figure}
\plotone{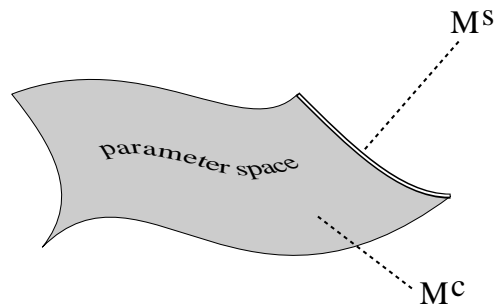}
\figcaption{Schematic representation of the simple and complex model parameter
spaces.
\label{arabadjis_f2}}
\end{figure}

Each model is fit to a data set consisting of two spectra, one from
the outer annulus and one from the inner annulus.  Because ${\sf M^s}$ lies on a
boundary of the parameter space of ${\sf M^c}$ (i.e.\ the normalization of the
second core component goes to zero; Figure~\ref{arabadjis_f2}), the standard
$F$ test cannot be used \citep{protassov}.  We must instead construct an
{\it empircal} $F$ distribution, using MCMC sampling (sketched below), to which
we apply the $F$ test.  The $F$ statistic,

\begin{equation}
F = \frac{\chi^2({\sf M^s|D}) -
\chi^2({\sf M^c|D})}{\chi^2({\sf M^s|D})/\nu({\sf M^s})} \, ,
\label{eq15}
\end{equation}

\noindent
is a measure of the improvement in a fit when the simple model is replaced by
the complex model.  The $F$ distribution of a large number of data realizations
allows us to quantitatively compare these two models by comparing the $F$
distribution to the $F$ value of the original data set.

Given the (known) probability distribution function $P({\bf x})$, where
${\bf x}$ is the vector of model parameters, we construct an empirical $F$
distribution using MCMC sampling and perform an $F$ test to determine the
significance of a second emission component in the core.  The entire procedure
is as follows:

\begin{verse}
                                                                                
1. Model the real data set ${\sf D}_0$ with $\sf M^s$ using XSPEC; call the
best-fit parameters ${\bf x}_0^{\sf s}$.
                                                                                
2. Use XSPEC to calculate $P({\sf D}_0|{\bf x})$ (i.e., the likelihood).
                                                                                
3. Use Bayes' Theorem to calculate $P({\bf x}|{\sf D}_0)$.

4. Create a large sample of model parameters ${\bf x}_i^{\sf s}$ using a random
walk through through the parameter space, rejecting each new position which
does not meet an acceptance criterion based upon $P({\bf x}|{\sf D}_0)$.
This is the Metropolis algorithm form of the MCMC technique; see \citet{neal}.
(We also discard all unphysical excursions in parameter space, i.e. where $T<0$
or $\rho<0$.)  For each ${\bf x}_i^{\sf s}$, compute a fake data set
${\sf D}_i$, including instrumental effects of the {\it Chandra} telescope and
detectors, as well as counting statistics.
                                                                                
5. Fit each $\sf M^s$ and $\sf M^c$ to each ${\sf D}_i$.

6. For each pair of models tabulate its $F$ value given by Equation~\ref{eq15}.
                                                                                
7. Bin up the set of $F$ values, creating an unnormalized histogram , and
compare to the $F$ value of the original data.
                                                                                
\end{verse}

In many applications of MCMC sampling one pays special attention to the finite
``burn-in'' period during which the Markov chain equilibrates.  The length of
the burn-in phase depends upon the sensibility of the starting point, and the
appropriateness of the scale chosen for the proposal probability distribution
step.  This is not a consideration in our case because we start each MCMC
sample at the (already known) peak of the probability distribution function
$P({\bf x})$.

In practice this recipe is computationally intensive, not because of any
features of the MCMC sampling {\it per se}, but because each of the faked
spectra must be modelled twice.  For a sample size of 1000 simulations,
XSPEC must simulate 1000 spectra and calculate 2002 sets of best-fit
values for the model parameters (including the original data).  This fact
leads us to simplify the method.  First, in order to reduce the modelling
time, we have adopted a simplified core-halo geometry.  In this scheme the
``core'' is represented by a single shell (in this case a sphere), while the
halo is represented by another shell.  Thus $\sf M^s$ contains four parameters,
the temperature and density of each of the two shells, while $\sf M^c$ contains
six, the additional two parameters representing the temperature and density of
a second cospatial emission component in the core.  This simplification also
greatly improves the numerical stability of the fitting procedure.  The
algorithm (steps 1-6 above) is implemented in a Tcl script run within XSPEC.
                                                                                
Once we have completed step 7 we can distinguish between the models.  The
location of the $F$ value of the data within this empirical $F$ distribution
contains information regarding the relative merit of $\sf M^s$ and $\sf M^c$.
We define the significance $S$ of the distribution as

\begin{equation}
\label{eq21}
s = \frac{\int_{0}^{F_{data}} N(F) \, dF}{\int_{0}^{\infty} N(F) \, dF}
\end{equation}
                                                                                
\noindent The signficance $S = 1 - P_f$, where $P_f$ is the probability that
the simple model constitutes the better description, and that the $F$ value of
the data is this large strictly by chance.  Thus, for a one-parameter model,
$S = 0.68$, 0.90, and 0.99 may be interpreted as 1, 2 and 3$\sigma$ detections
of the additional component.  When discussing the presence of a second emission
component in the X-ray data, we adopt 99\% as a threshold significance.

\section{Application to Chandra Spectra and Clusters}
\label{arabadjis:application}

We apply this method to 12 clusters observed with Chandra.  All clusters in the
sample are fairly round and, with the exception of CL0024, seem to be fairly
spherical with no significant amounts of substructure.  Ten of these
clusters are generally agreed to contain ``cooling flows'' in their centers,
while two of them, CL0024 and A2104, do not.  (\citet{allen} derives an upper
limit to the cooling rate in A2104.)  In each cluster, the core size is defined
as the radial extent containing roughly 3000-4000 source photons.  Except for
Hydra A, all of the clusters in the sample admit a second emission component in
this region.  (In Hydra A, the best-fit temperature of the second component
equals that of the first, rendering an MCMC $F$-test irrelevant.)

The resulting F distributions for the entire sample are shown in
Figure~\ref{arabadjis_f3}.  Statistics for each cluster are shown below.  We
list list the cluster, the pre-Chandra/XMM cooling rates from the literature, a
reference for this value, and the multiphase plasma MCMC signficance.

\vspace{10pt}

\small {\sc Table} 1.-- The 12 Chandra clusters in the sample. \normalsize

\vspace{10pt}

\begin{tabular}{cclclcc}
           & \hspace{5pt} & \hspace{15pt} $dM/dt$ & \hspace{10pt} & \hspace{10pt} & & $S$ \\
  cluster  &  & \hspace{5pt} (M$_\odot$ y$^{-1}$)      & & \hspace{8pt} reference & & ($N=1000$) \\
A1689  &  & \hspace{5pt} $118  \, \pm ^{375}_{118}$ & & White (2000) & &  34.2\% \\
A1795  &  & \hspace{5pt} $453  \, \pm ^{86}_ {90} $ & & White (2000) & &  82.3\% \\
A1835  &  & \hspace{5pt} $683  \, \pm ^{677}_{677}$ & & White (2000) & &  48.3\% \\
A1942  &  & \hspace{5pt} $817  \, \pm ^{118}_{741}$ & & White (2000) & &  17.6\% \\
A2029  &  & \hspace{5pt} $547  \, \pm ^{72}_ {81} $ & & Allen \hspace{1pt} (2000) & &  99.8\% \\
A2104  &  & \hspace{15pt} $0    \, \pm ^{94}_ {0}  $ & & White (2000) & &  66.1\% \\
A2204  &  & \hspace{5pt} $984  \, \pm ^{583}_{653}$ & & White (2000) & &  99.9\% \\
CL0024$^*$ &  & \hspace{15pt} --$   \,              $ & & \hspace{20pt} -- & & 77.0\% \\
HydraA &  & \hspace{5pt} $264  \, \pm ^{81}_ {60} $ & & Allen \hspace{1pt} (2000) & &   --  \\
MS1358 &  & \hspace{5pt} $691  \, \pm ^{348}_{287}$ & & Allen \hspace{1pt} (2000) & &  63.4\% \\
MS2137 &  & \hspace{0pt} $1467 \, \pm ^{880}_{726}$ & & Allen \hspace{1pt} (2000) & &   27.1\% \\
ZW3146 &  & \hspace{0pt} $2228 \, \pm ^{357}_{636}$ & & Allen \hspace{1pt} (2000) & &  98.9\% \\
\end{tabular}

$^*$ \small There are no measurements (nor upper limits) of \\
\vspace*{-2pt}\hspace*{20pt} cooling flow plasma in CL0024 in the literature.
\normalsize

\vspace{10pt}

\section{Discussion}
\label{arabadjis:discussion}

Of the 10 cooling flow clusters in the sample, only three of them -- A2029,
A2204, and ZW3146 -- show evidence for multiphase gas in the core, if we
adopt $S=99\%$ as our significance threshold.  Perhaps more surprising is the
fact that the two clusters with the largest pre-Chandra/XMM mass deposition
rates -- MS2137 and ZW3146 -- display such dramatically different evidence for
the existence of  multiphase plasma.  Both were suspected of harboring cooling
flows with rates in excess of 1000 M$_{\odot}$ y$^{-1}$, and yet MS2137 shows
no evidence of multiphase gas.  (As expected, the two non-cooling flow clusters
also show no evidence for a second core emission component.)  It could be that,
in the absence of recent core merging events, the equilibrium state of the
plasma is uniphase, and that the merger and accretion of smaller (and cooler)
substructures during the continuing assembly of the cluster are responsible for
some clusters showing evidence of multiphase cores.  If this were the case,
one might look for some evidence for a merging event in MS2137 which has
occured within a cooling time.  (As yet we have not found none.)  Regardless,
it seems that cooling flow cluster cores are an inhomogeneous class.

\begin{figure}
\plotone{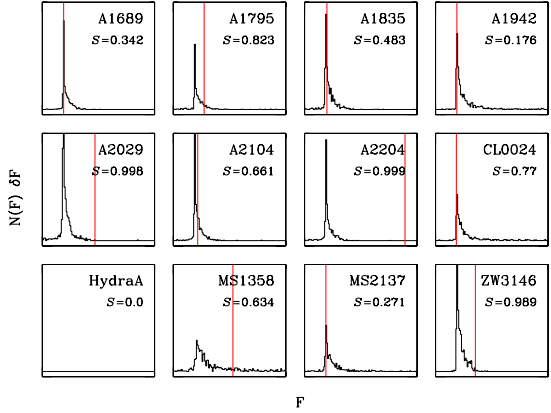}
\figcaption{Empirical F distributions for the 12 clusters in this study.  Note
that Hydra A does {\it not} admit a second core component whose temperature
differs from the first.
\label{arabadjis_f3}}
\end{figure}





\end{document}